\documentclass{article}

\usepackage{PRIMEarxiv}

\usepackage[utf8]{inputenc} % allow utf-8 input
\usepackage[T1]{fontenc}    % use 8-bit T1 fonts
\usepackage{hyperref}       % hyperlinks
\usepackage{url}            % simple URL typesetting
\usepackage{booktabs}       % professional-quality tables
\usepackage{amsfonts}       % blackboard math symbols
\usepackage{nicefrac}       % compact symbols for 1/2, etc.
\usepackage{microtype}      % microtypography
\usepackage{lipsum}
\usepackage{fancyhdr}       % header
\usepackage{graphicx} 
\usepackage{xcolor} % graphics
\graphicspath{{media/}}     % organize your images and other figures under media/ folder

\title{Breaking down the relationship between academic impact and scientific disruption
\thanks{\textit{\underline{Citation}}: 
\textbf{Authors. Title. Pages.... DOI:000000/11111.}} 
}

\author{
  Mingtang Li\\
  Department of Computer Science \\
  University College London \\
  London (UK)\\
  \texttt{mingtang.li.20@ucl.ac.uk} \\
   \And
  Giacomo Livan \\
  Dipartimento di Fisica \\
  Universit\`a degli Studi di Pavia \\
  Pavia (Italy); \\
  Department of Computer Science \\
  University College London \\
  London (UK)\\
  \texttt{giacomo.livan@unipv.it} \\
   \AND
  Simone Righi \\
  Department of Economics \\
  University Ca' Foscari of Venice \\
  Venezia (Italy) \\
  \texttt{simone.righi@unive.it} \\
}

\begin{document}
\maketitle

\begin{abstract}
We examine the tension between academic impact -- the volume of citations received by publications -- and scientific disruption. Intuitively, one would expect disruptive scientific work to be rewarded by high volumes of citations and, symmetrically, impactful work to also be disruptive. A number of recent studies have instead shown that such intuition is often at odds with reality. In this paper, we break down the relationship between impact and disruption with a detailed correlation analysis in two large data sets of publications in Computer Science and Physics. We find that highly disruptive papers tend to be cited at higher rates than average. Contrastingly, the opposite is not true, as we do not find highly impactful papers to be particularly disruptive. Notably, these results qualitatively hold even within individual scientific careers, as we find that -- on average -- an author's most disruptive work tends to be well cited, whereas their most cited work does not tend to be disruptive. We discuss the implications of our findings in the context of academic evaluation systems, and show how they can contribute to reconcile seemingly contradictory results in the literature.
\end{abstract}

% keywords can be removed
\keywords{scientific impact \and scientific disruption \and scientific careers}

\section{Introduction}\label{sec1}

In an increasingly competitive academic environment, the performance of researchers is constantly monitored, quantified, and ranked in a variety of dimensions. Some of these can be measured rather objectively (e.g., productivity, ability to attract funding, etc.~\cite{Hicks2012,Abramo2014,Cattaneo2016}), while others are more elusive, such as the ability to innovate and/or to produce impactful research~\cite{Redner1998,Buchanan2015}. Conventionally, these dimensions are often measured as a function of the citations received by published work~\cite{Moed2005,Waltman2016}, either via simple citation counts or via more sophisticated bibliometric indicators, such as the well-known $h$-index~\cite{Hirsch2005}, $g$-index~\cite{Egghe2006}, or indicators of an author's  performance relative to their field (see, e.g.,~\cite{Radicchi2008}). These indicators reflect the extent to which research outputs are recognized by the scientific community. However, they also play an increasingly pervasive role in research evaluation systems, as they influence research rankings, grant attributions, tenure and promotion decisions~\cite{Garfield1979,Holden2005,Carlsson2009,Clavijo2013,Owens2013}.

Given the significance of citation metrics in academic evaluation, a growing number of studies have been devoted to investigating the factors shaping the number of citations received by a paper. Among these factors, interdisciplinarity has a considerable influence on scientific impact~\cite{Wang2017,Fortunato2018}. Indeed, it has been found that `long-distance' interdisciplinary research on average attracts citations at higher rates~\cite{Larivière2015}, but there exists an interdisciplinarity `tipping point' beyond which highly interdisciplinary publications tend to have lower impact~\cite{Larivière2010,sun2021interdisciplinary}. In fact, papers that are more likely to be highly cited tend to draw heavily from conventional combinations of existing research while still integrating unusual combinations~\cite{Uzzi2013}. The accumulation of citations and its determinants are also studied from the viewpoint of authors and their career progression. For instance, it is well known that scientific careers are characterized by the so called `random impact rule', i.e., each paper within a authors's publication sequence has the same likelihood of becoming their most-cited work~\cite{Sinatra2016}. Nevertheless, authors can experience `hot streak' periods during which they produce a series of high-impact papers~\cite{Liu2018,Liu2021}.

Citation-based bibliometric indicators have been increasingly scrutinized by the academic community and have become somewhat controversial~\cite{Garfield1979,Phelan1999,Abramo2011,Werner2015,Baccini2019}. One of the major concerns is that such indicators --- and citations in general --- are not a comprehensive proxy of scientific innovation~\cite{Moed2005,Aksnes2019,Fanelli2009,Fang2012}. To better quantify the innovativeness of scientific outputs, the CD index, also known as the disruption score, has been put forward~\cite{Funk2017,Wu2019}. This indicator has been applied as a measure of innovation in a variety of studies~\cite{Wu2019,Chu2021,Park2023,Wei2023}, and it has been proven to be effective at distinguishing between disruptive and developmental works. Despite its surging popularity, the disruption score has been criticized for being temporally biased and easily distorted by citation inflation~\cite{Petersen2023}. We anticipate that in this paper we will adopt a variant of the disruption score to mitigate these potential biases (see Methods).

A number of studies have leveraged the disruption score to explore scientific dynamics that cannot be explained by citations or impact. For instance, papers with a larger number of authors are more likely to be cited~\cite{Tahamtan2016}. However, papers authored by large teams tend to be developmental, disruptive research tends to be produced by smaller teams~\cite{Wu2019}. A recent study investigated the relationship between productivity, innovation, and impact, showing that the authors typically produce more innovative work during periods of low productivity. Conversely, high-impact publications tend to be produced during stretches of high productivity~\cite{Li2023}. Another very recent paper found that papers and patents are becoming less disruptive over time~\cite{Park2023}.

The above findings show that scientific impact and innovation exhibit rather different patterns, almost to the point that they should be treated as two distinct concepts~\cite{Aksnes2019}. Yet, the combination of such two concepts has also been shown to be effective, e.g., as a way to identify revolutionary scientific contributions. In fact, Nobel Prize-winning papers generally obtain more citations and achieve higher disruption scores~\cite{Wei2023}. However, such a result seems to contradict the findings by Zeng~\emph{et al.} that disruptive papers in science are losing impact~\cite{Zeng2023}.

Motivated by these observations, in this paper we seek to fully explore the relationship between scientific impact and innovation. We begin by breaking down the correlation between disruption scores and citations across each percentile of the top disruptive papers. Then we uncover the full picture of the relationship between disruption scores and citations by investigating whether the most cited papers in a field are also disruptive. Finally, we extend our paper-level findings to the context of career analysis, showing that the relationship between disruption scores and citations also holds at the level of entire careers.

\section{Results}\label{sec2}

We collect papers published between 1986 and 2015 in Computer Science and and Physics from the AMiner citation network dataset (version 12) and the Web of Science database, respectively (see Methods). We associate a disruption score to each paper, which characterizes a paper as more disruptive when ensuing publications in the same field cite such a paper at a higher rate than the publications in its bibliography (see Methods). We quantify scientific impact as the citations accumulated over the first five years after publication, which is a customary proxy in the literature~\cite{Park2023,Zeng2023,Sun2023}. Overall, our analysis comprises 898,624 papers in Computer Science and 1,236,016 papers in Physics.

\subsection{A detailed breakdown on the correlations between disruptions and citations}\label{sec21}

%We begin by further exploring the findings proposed by Zeng~\emph{et al.} on the relationship between disruptions and citations~\cite{Zeng2023}. 

We begin our analysis with a detailed breakdown of the correlation between scientific disruption and impact. Namely, we  rank all the papers in our dataset based on their disruption scores and their impact. In the following, we shall refer to the rankings computed via disruption scores and citations as the `disruption rank' and the `impact rank', respectively. We select all papers in the top 1\% of the disruption rank, and compute the Kendall correlation coefficient with their positions in the impact rank. We then repeat this process for papers in the top 2\%, top 3\%, etc, until all papers in Computer Science (898,624 papers in total, 8,986 papers in each percentile) and Physics (1,236,016 papers in total, 12,360 papers in each percentile) have been included. 

\begin{figure}[h!]
\centering
\includegraphics[width=0.8\textwidth]{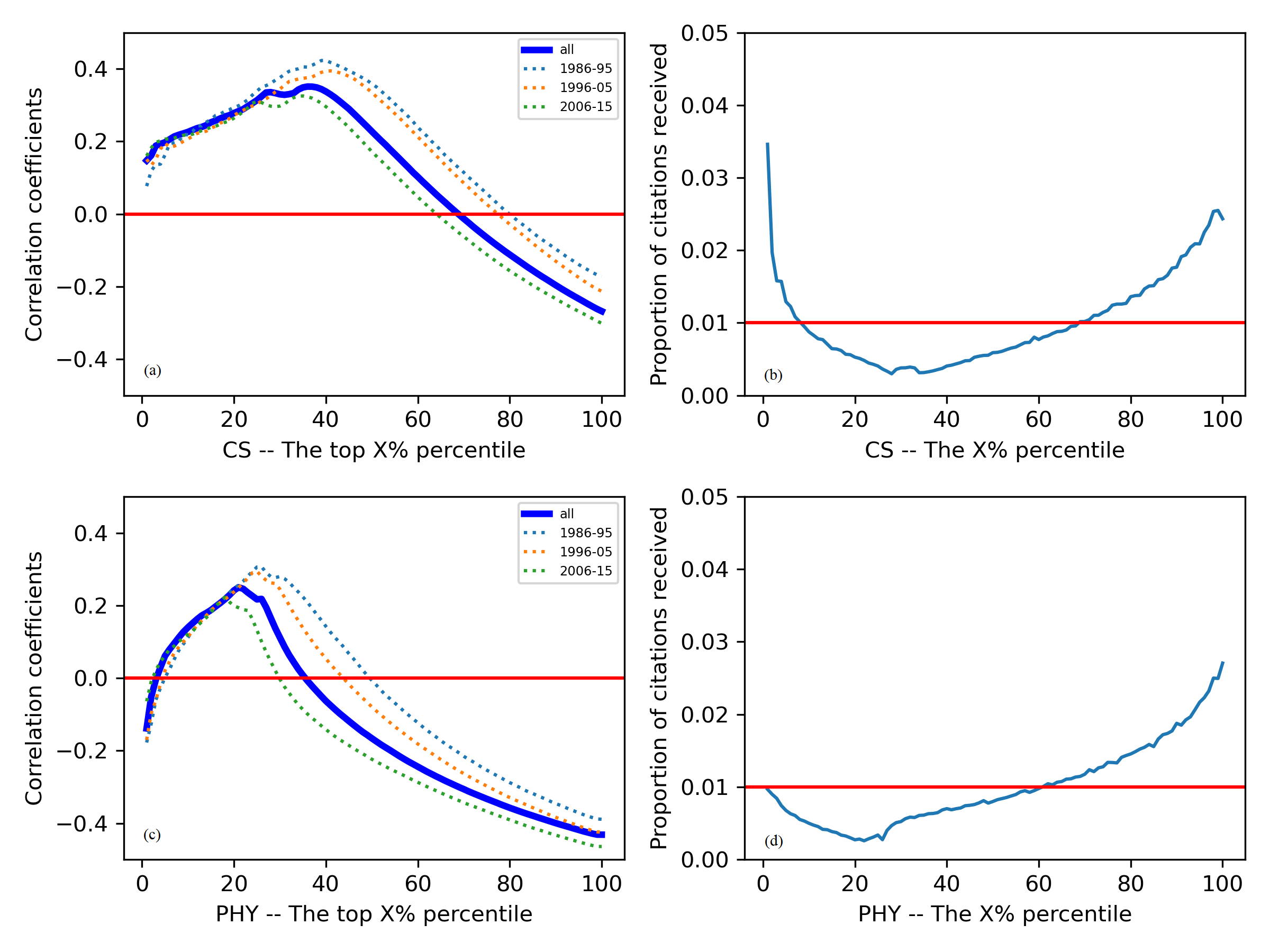}
\caption{(a) Correlation coefficient across different percentiles of disruptive papers in Computer Science. The thick blue curve is derived from all the papers in our datasets, while dashed lines represent the correlation coefficients corresponding to the 1986-1995, 1996-2005, and 2006-2015 groups. (b) The proportion of citations received by each percentile of Computer Science papers. The correlation pattern observed in (a) can be explained by the proportion of citations received as shown in (b). (c) and (d) are the equivalent versions of the correlation trajectory and the proportion of received citations in Physics. They can be interpreted in a similar way to (a) and (b).}
\label{fig:R1Fig}
\end{figure}

In Fig. ~\ref{fig:R1Fig} (a)-(c) we plot the aforementioned correlation coefficients as a function of percentiles of the disruption distribution in Computer Science and Physics. We observe a positive correlation coefficient between disruption and impact for papers in the top percentiles of the disruption distribution. Such correlation increases as we incorporate more percentiles into our analysis, reaching a peak value around the top 25th percentile, then declining to negative values. To explain such a pattern, in Fig.~\ref{fig:R1Fig} (b)-(d) we report the proportion of citations received by papers in each percentile of the disruption distribution. We find that the papers receiving the lowest share of citations are those around the 25th percentile, i.e., where we observe the peak in correlation between disruption and impact. After that, less disruptive papers progressively become more cited, which causes the correlation coefficient to decrease. Eventually, the correlation coefficient becomes negative when we consider a large enough portion of papers in our dataset, which supports the result by Zeng~\emph{et al.} on the negative correlation between disruption and impact.

%We find that within the top 1\%-25\% percentile of the most disruptive papers, those having relatively lower disruption levels also receive fewer proportion of citations. Such papers rank lower in both disruption and citation ranks, thus resulting in a higher correlation coefficient. After the peak correlations, the papers we consider become progressively less disruptive yet attracting a higher proportion of citations, which in turn causes the correlation coefficient to diminish. Eventually, the correlation trajectory ends up with a negative value when we consider all papers in our datasets, which supports Zeng~\emph{et al.}'s findings on the negative correlation between disruptions and citations for academic papers~\cite{Zeng2023}.

Fig.~\ref{fig:R1Fig} (b)-(d) also show that the most disruptive papers are quite well recognized, as evidenced by the relatively higher proportion of citations received by the most disruptive papers in both Computer Science and Physics, although with remarkable differences. In fact, highly disruptive papers in Computer Science are cited at a rate which is much higher than one would expect from a random baseline (i.e., all percentiles receiving a 1\% share of all citations). The same cannot be said for Physics, where the most disruptive papers are cited at a rate which is slighlty lower than the random baseline. These differences are responsible for the positive (negative) correlation between disruption and impact observed in the top percentiles of the disruption distribution in Computer Science (Physics).

%This result demonstrates that the most disruptive papers do not have their disruption scores distorted by receiving only a small number of citations. Instead, such disruptive papers obtain notable scientific impact and truly eclipse the attention of future works to previous publications. Still, the recognition of the top disruptive papers exhibits disciplinary characteristics. While the top 1\% disruptive papers in Computer Science have a positive correlation with their positions in the citation rank, the correlation coefficient between disruptions and citations for the most disruptive papers in Physics is negative.

We test the robustness of the aforementioned results in three different ways. First, we split all  papers in our dataset into three groups based on their publication year, namely 1986-1995, 1996-2005, and 2006-2015, and then repeat the above experiment for each group. The aim of this test is to illustrate that our results are robust over different periods. As can be seen in Fig.~\ref{fig:R1Fig} (a) and (c), we find consistent patterns across the three groups. Second, we standardize the disruption score (see Methods) of each paper to account for the fact that papers tend to become less disruptive over time~\cite{Park2023}. We perform the same experiment with the standardized disruption scores, obtaining consistent results across the two disciplines (see Appendix Fig.~\ref{fig:R1Supp1}). Third, we run the same experiments with a null model created by reshuffling the 5 years of accumulated citations received by each paper while keeping their disruptions intact. By reshuffling citations, we randomize the position of top disruptive papers in the impact rank, thus the new correlation coefficients are calculated under the null model. We find that the correlation patterns cannot be explained by the null model, and the correlation coefficients across different percentiles of disruptive papers are around 0 (see Appendix Fig.~\ref{fig:R1Supp2}). 

\subsection{Most-cited papers are relatively less disruptive}\label{sec22}

After examining the relationship between disruption and impact across various percentiles of the disruption distribution, we now explore the relationship from the opposite perspective, i.e, by analysing different percentiles of the impact distribution. Similar to the procedure described in the previous section, we choose the top 1\% most cited papers in the impact rank and identify their respective positions in the disruption rank. We calculate the correlation coefficient between these two position vectors and repeat the procedure for the top 2\%, 3\%, up to all the papers in both disciplines.

\begin{figure}[h!]
\centering
\includegraphics[width=0.8\textwidth]{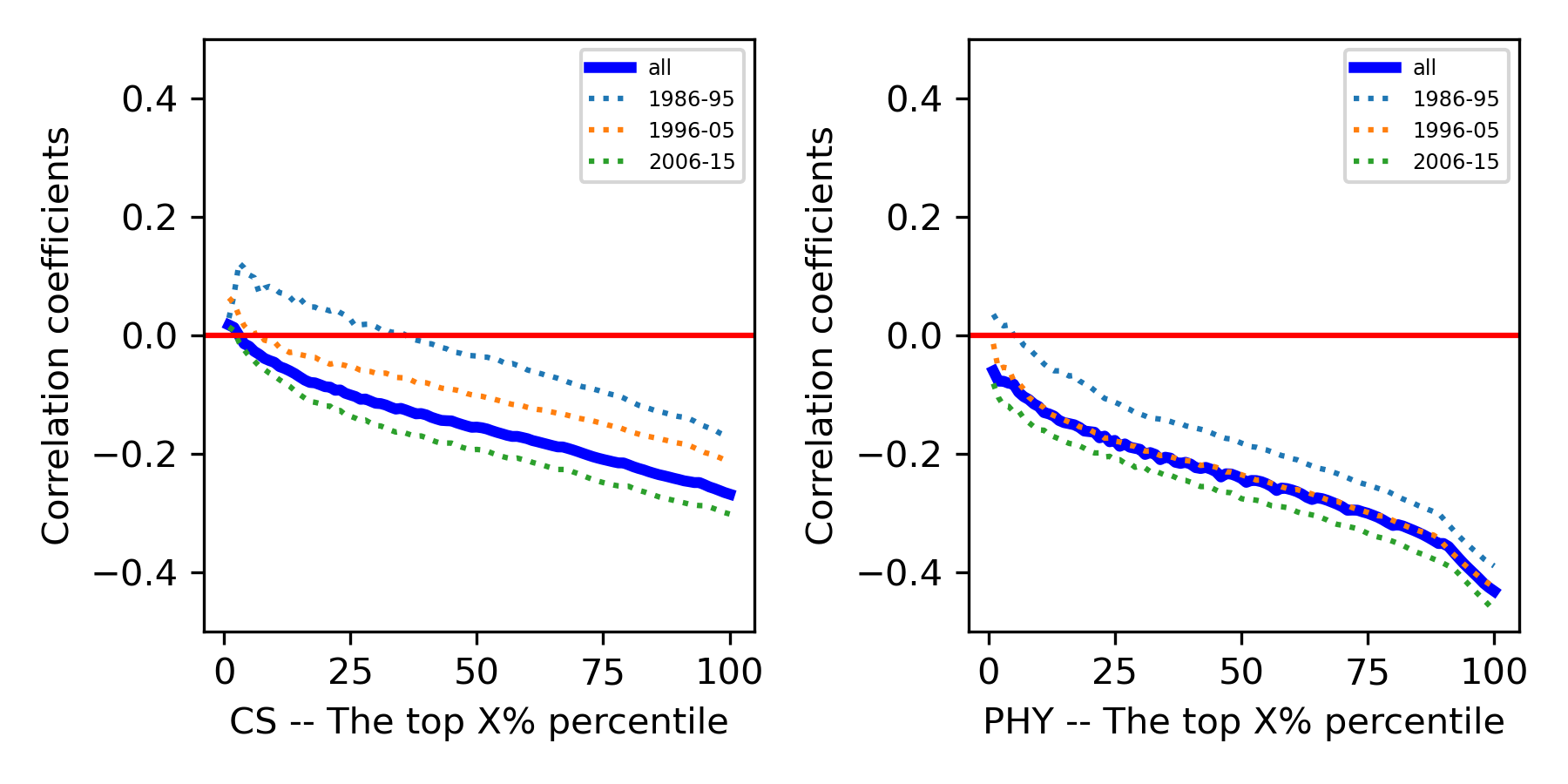}
\caption{Correlation coefficient across different percentiles of the most-cited papers in Computer Science (left) and Physics (right). The patterns of correlation coefficients are very similar in both disciplines, which indicates that the most-cited papers tend to be less disruptive.}
\label{fig:R2Fig}
\end{figure}

As shown in Fig.~\ref{fig:R2Fig}, a negative correlation coefficient is apparent across most percentiles of the impact rank in both Computer Science and Physics. The negative correlation strengthens as we incorporate more percentiles into our analysis. Such a pattern indicates that the most-cited papers tend to be less disruptive, and vice versa. Moreover, in both disciplines, the correlation coefficients are generally higher for papers published between 1986 and 1995. In particular, we can find a positive correlation coefficient in the top 1\%-30\% of the most-cited Computer Science papers. By contrast, for papers published in more recent decades (1996-2005, 2006-2015), the correlation trajectory is not only negative over the entire distribution, but the negative correlation becomes even more pronounced. To further corroborate these results, we perform the same experiment with the standardized disruption scores and with the null model, through the same methods described in the previous section. Our results are valid under both robustness tests (see Appendix Fig.~\ref{fig:R2Supp1}).

\subsection{The relationship between disruptions and citations in scientific careers}\label{sec23}

Given the aforementioned results, a natural extension of our research is to investigate the relationship between disruption and impact in scientific careers. To construct an author-centred dataset, we first match each paper in our datasets with its respective authors, and then identify long-lived researchers with an active publication record. Specifically, we only include in our analysis authors who started their careers between 1980 and 2000 and had an academic career of at least 20 years. Among these authors, we retain only those who published more than 10 papers, with a publication frequency of at least one paper every 5 years (in line with~\cite{Li2019}). Based on these selection criteria, we are left with 27,598 Computer Scientists and 34,527 Physicists (see Methods).

\begin{figure}[h!]
\centering
\includegraphics[width=0.8\textwidth]{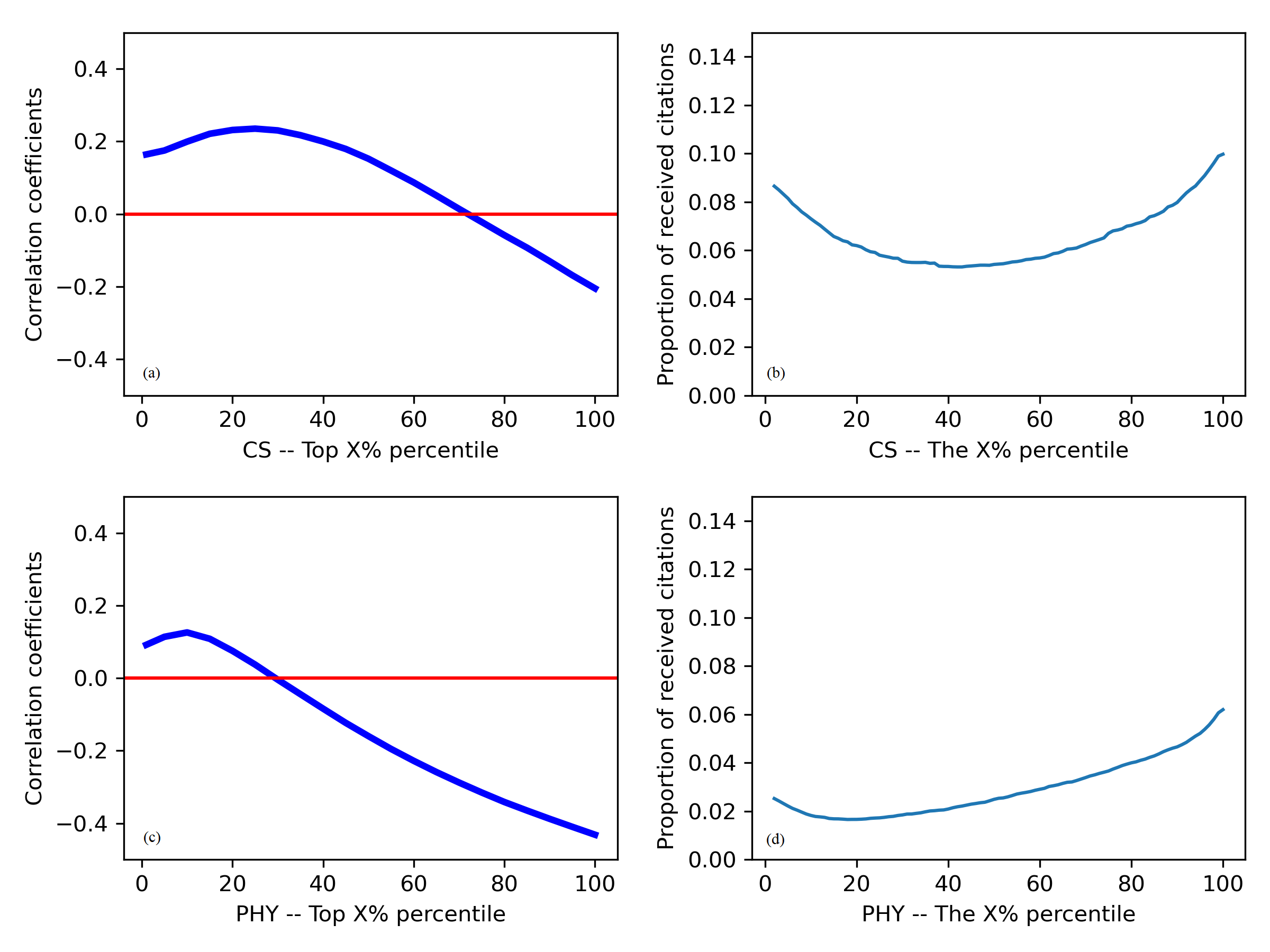}
\caption{(a) The mean value of correlation coefficient across different percentiles of disruptive papers in the careers of Computer Scientists. (b) The average value of the proportion of citations received by papers at each percentile in the publication sequence of Computer Science researchers. Again, the correlation pattern observed in (a) can be explained by the proportion of citations received as depicted in (b). (c) and (d) are the equivalent version of (a) and (b) in Physics. We can observe that in both disciplines, our paper-level results are consistent in the career-level analysis.}
\label{fig:R3Fig1}
\end{figure}

We first seek to generalize the findings concerning the disruptive papers to our pool of researchers. Following the method described in~\ref{sec21}, we start by creating a disruption rank and an impact rank for papers published during a scientific career. We then locate the top 1\% disruptive papers within both disruption and citation rank, and compute the correlation coefficient between these two sets of ranks. For this analysis, we do not calculate the correlation coefficient for each top percentile of papers because doing so can produce an excessive number of repeated values due to the relatively small amount of publications in the publication sequence of an author. Instead we repeat such a process exclusively for the top 1\%, 5\%, 10\%, 15\%, and so on. Finally, we collate the correlation coefficients for the same top percentile across all authors in Computer Science and Physics, and plot the mean values of each top percentile in Fig.~\ref{fig:R3Fig1} (a) and (c). 

Similar to section~\ref{sec21}, we also show the proportion of citations received by each percentile of papers in disruption distributions at the author level Fig.~\ref{fig:R3Fig1} (b) and (d). We observe that the overall trends in panels (b) and (d) are comparable to the trends in the paper-level results. It is noted that the curves in (b) and (d) achieve higher values compared to the results in the paper datasets. This happens because the same papers can fall within different percentiles when considering less prolific authors. When we restrict our scope of investigation to researchers who have more than 100 publications, i.e., no overlaps between percentiles, our results are very much similar to the paper-level results (see Appendix Fig.~\ref{fig:R3Supp3}).

As can be seen in Fig.~\ref{fig:R3Fig1}, the findings we obtain here are fairly similar to those we observe at the paper level. Specifically, the most disruptive papers in the careers of Computer Scientists and Physicists still attract a relatively high proportion of citations. The correlation trajectories in scientific careers also display a pattern of initial increase followed by a decrease, and such a trajectory can also be explained by the proportion of citations received by each percentile of papers published in a career. Furthermore, we observe a negative correlation coefficient when considering all papers in a career, indicating that the overall negative relationship between disruption and impact persists at the career level. The only significant difference between our findings in academic publications and scientific careers is that the correlation coefficients for the most disruptive papers are now positive in both disciplines. This suggests that the most disruptive papers within a career are well rewarded in terms of impact by their respective scientific communities.

\begin{figure}[h!]
\centering
\includegraphics[width=0.8\textwidth]{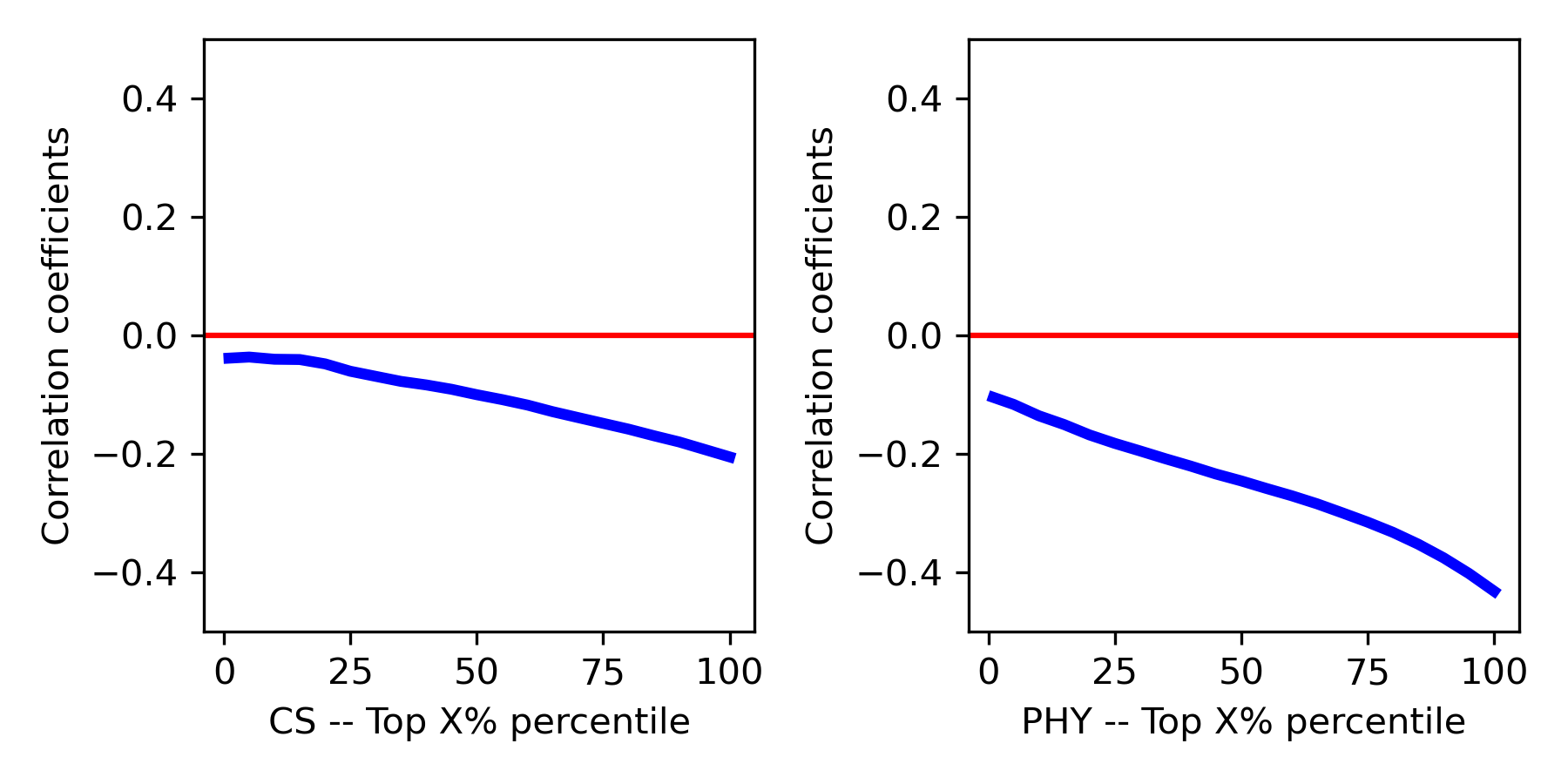}
\caption{Correlation coefficient across different percentiles of the most-cited papers in the careers of Computer Science (left) and Physics (right). It can be seen that our paper-level results hold true in scientific careers.}
\label{fig:R3Fig2}
\end{figure}

We then expand our results regarding the correlations for the most-cited papers to the context of scientific careers. To achieve this, we follow the steps outlined in~\ref{sec22} and compute rank-rank correlations over an increasing number of percentiles of the impact distributions obtained at the career level. The results are illustrated in Fig.~\ref{fig:R3Fig2}. In line with our results at the paper level, we can still observe negative correlation coefficients across each impact percentile in the careers of both Computer Scientists and Physicists. This reinforces our conclusion that the most-cited papers tend to be less disruptive.

In order to further substantiate these findings, we repeat the aforementioned experiments using the standardized disruption score and obtain consistent results (see Appendix Fig.~\ref{fig:R3Supp1} and Fig.~\ref{fig:R3Supp4}). Moreover, we construct two null models in a similar manner to the previous experiments by reshuffling the disruption score and the 5-year accumulated citations in each author's publication sequence. We then reapply the career-level analysis utilizing these null models, and find that our conclusions cannot be explained by these null models (see Appendix Fig.~\ref{fig:R3Supp2} and Fig.~\ref{fig:R3Supp4}). Based on all these results, we believe that our paper-level conclusions regarding the relationship between disruptions and citations still hold in scientific careers.

\section{Discussion}\label{sec3}

We examined the relationship between scientific innovation and impact, measured in terms of disruption scores and citations, respectively. Our aim is to fully capture the relationship between such two dimensions through two main research questions, namely (1) are disruptive papers highly cited?; and (2) are high-impact papers disruptive? 

To answer the first question, we analyzed the correlation coefficients between disruption scores and citations across different percentiles of papers ranked by their disruption scores. In both Computer Science and Physics, we find that the correlation varies when we observe different samples of disruptive papers, and that the variations in the correlation coefficients can be explained by the proportion of the citations received by each percentile of papers. Our results reconcile the seemingly contradictory conclusions between Wei~\emph{et al.}~\cite{Wei2023} and Zeng~\emph{et al.}~\cite{Zeng2023}. Specifically, papers with higher levels of disruption exhibit a positive correlation between disruption scores and citations. This pattern is consistent with the finding, e.g., that Nobel Prize-winning papers typically receive more citations and are characterized by higher disruption scores~\cite{Wei2023}. However, as we incorporate more percentiles in the disruption score distribution, the correlation coefficient gradually shifts from positive to negative values, and ends up with a negative correlation when we include most of the papers in our analysis, in line with findings by Zeng~\emph{et al.}~\cite{Zeng2023}. Concerning the second question, we find a negative correlation between disruption scores and citations in both disciplines, which suggests that the most-cited papers tend to be less disruptive. Moreover, we observe that such a negative correlation intensifies over time.

Having determined the relationship between disruption scores and impact at the level of academic publications, we then expand our results to the careers of Computer Scientists and Physicists, concluding that the aforementioned  results remain equivalent at the level of careers. Our results suggest that there are two strategies researchers might adopt to maximize their citation rates. The first strategy aims to publish truly disruptive papers. This strategy is beneficial to the development of science as a whole but requires researchers to accumulate research experience, go through periods of focus and low productivity~\cite{Li2023}, and undertake the risk of receiving only a limited number of citations. The second strategy is to produce papers that attract a large number of citations. Such a strategy favors the career progression of individual researchers. However, it may also incentivize researchers to focus excessively on popular research topics and incremental work, which can be detrimental to the overall diversity and innovation of scientific research~\cite{Livan2019}.

A common criticism of the disruption metric is that the score of a paper can be distorted upward by receiving only a small number of citations~\cite{Park2023}, i.e., high scores do not indicate high research quality but might simply reward papers that are less appreciated by citations. However, our results show that the top 1\% of disruptive papers not only achieve high disruption score levels but also attract a high proportion of citations. These findings are consistent in both Computer Science and Physics, and apply to both paper-level and career-level analysis. Therefore, such a criticism does not apply to papers with very high disruption scores.

The evaluation of research outputs is often based on bibliometric indicators of scientific impact~\cite{moher2018assessing}. Our study reveals that when research assessment relies excessively on citations, papers that stand out in this regime tend to be less disruptive. Similarly, if evaluations were to be purely based on disruption scores, some high-scoring papers may also exhibit limited scientific impact. A more effective approach would integrate both innovation and impact as complementary dimensions. Such an approach would enable us to identify papers that are both disruptive and impactful. Papers that excel under those criteria are typically recognized as work of very high quality~\cite{Wei2023}. Therefore, we advocate that scientific evaluation should be carried out through a comprehensive analysis of publications~\cite{Aksnes2019,Hicks2015}.

\section{Methods}\label{sec4}
\subsection{Data}\label{sec41}

We collect publication and citation data pertinent for Computer Science from the AMiner citation network dataset (version 12). The AMiner dataset extracts papers published between 1960s to 2020 from DBLP, ACM, MAG, and other sources in Computer Science~\cite{Tang:08KDD}, and it records a total of 4,894,081 papers and 45,564,149 citation relationships. The AMiner dataset has been utilized in a variety of bibliometric studies~\cite{Arif2014,zeng2019increasing,Anil2020,Li2023}.

For publications in Physics, we retrieve data from the Web of Science (WOS) database. We extract the papers published by long-lived researchers who maintain an active publication record, along with the citation network related to their publications. In total, we collect 1,619,039 papers and 12,621,175 citation relationships from 1985 to 2020. It is important to note that the WOS database does not provide unique author identifiers. To link authors with their respective publications, we employ the method proposed by Caron~\emph{et al.} to disambiguate author names~\cite{caron2014large}. This method determines a similarity score between pairs of authors by considering various attributes, such as ORCID identifiers, names, affiliations, emails, coauthors, grant numbers, etc. If a pair of authors has a higher similarity score, they are more likely to be identified as the same person. The effectiveness of this method has been validated by a recent study with precision and recall scores higher than 90\%~\cite{tekles2020author}.

In our analysis, we only calculate disruption scores for papers published before 2016, thereby allowing papers in our pool to accumulate citations for at least 5 years. We set filtering criteria for researchers in line with~\cite{Li2019}, performing our career analysis on a total of 27,598 and 34,527 researchers in Computer Science and Physics, respectively.

\subsection{The disruption score}\label{sec42}

We employ the disruption score to quantify the disruption level of each paper in our datasets. The fundamental idea of the disruption score is that a highly disruptive publications can overshadow preceding papers in the same field. The subsequent papers are more likely to cite the disruptive work over the references listed in its bibliography. The disruption score is particularly useful in differentiating between disruptive and developmental pieces of work, and it has been validated using data from academic publications, patents, and software products~\cite{Funk2017,Wu2019,Park2023}.    

To be more specific, we create a citation network centered on a focal papers, combined with its references (preceding papers) and subsequent papers. The subsequent papers can be further categorized into three groups: papers citing only the focal paper, those citing both the focal paper and the references, and those citing only the references of the focal paper. Let us assume that the number of subsequent papers in the three groups are $n_i$, $n_j$ and $n_k$, respectively. Then the disruption score can be determined as

\begin{equation}
D = \frac{n_i - n_j}{n_i + n_j + n_k}
\end{equation}

where $n_i - n_j$ quantifies the extent to which the focal paper has eclipsed attention towards preceding papers, and $n_i + n_j + n_k$ represents the total number of subsequent papers within the citation network.

According to the above definition, the disruption score spans from -1 to 1. A positive score indicates that the focal paper draws more attention from subsequent publications than its references. By definition, such a focal paper is more disruptive. If a focal paper is disruptive enough, then its disruption score $D$ should be close to 1. Conversely, a negative score implies that the focal paper is more likely to be developmental. The focal paper exhibits an increasing degree of its developmental character as its disruption score approaches to -1. Overall, the disruption score not only allows us to quantify the disruption of each paper, but also to compare the disruption level among different publications.

We also note that the disruption score can be represented by an alternative formula given as

\begin{equation}
D = \frac{1}{n}\sum_{i=1}^{n}-2f_{i}b_{i} + f_{i}
\end{equation}

where $n$ denotes the total number of subsequent papers in the citation network, $i$ represents the collection of subsequent works that cite the focal paper and/or the focal paper's references, $f_{i} = 1$ if $i$ only cites the focal paper and $0$ otherwise, and $b_{i} = 1$ if $i$ cites any predecessors of the focal paper and $0$ otherwise.

These two expressions of the disruption score are mathematically equivalent. For the second expression, if a subsequent paper $i$ cites only the focal paper, i.e., belonging to the $n_i$ group, then $-2f_{i}b_{i} + f_{i} = 1$ as $-2f_{i}b_{i} = 0$. When a subsequent paper cites both the focal paper and its predecessors (within the $n_j$ group), then the value of $-2f_{i}b_{i} + f_{i}$ will be -1. If a subsequent paper belongs to the $n_k$ group, then $-2f_{i}b_{i} + f_{i}$ equals 0. By summing the $-2f_{i}b_{i} + f_{i}$ terms across all the subsequent papers in the citation network, the result equals to the difference between the number of $n_i$ papers and $n_j$ papers, which is the $n_i - n_j$ term in the first formula.

A major concern about the disruption metric is that the disruption level of a paper can be inflated by receiving a small number of citations. To solve this problem, we do not impose any restrictions on the period of investigation, and we include all the subsequent papers within the citation network to compute the disruption score. Nevertheless, this approach may cause other biases, such as the overall decline in papers' disruptions over time, the changing citation behaviours for papers published at different times, and the reduced comparability of the disruption metric across years. To mitigate these biases, we standardize the disruption score of each paper with respect to its year of publication. Specifically, we group papers by their respective publication years, and then standardize their disruption levels by incorporating the mean and standard deviation of that year’s distribution of disruption scores (i.e., transforming into z-scores). In this paper, our findings remain consistent in both the original and standardized disruption scores, which further corroborates the validity of our results.

\section*{Acknowledgments}
G.L. acknowledges support from a Leverhulme Trust research project grant (RPG-2021-282). SR gratefully acknowledges funding the Italian Minister of University and Research (MUR) under Research Projects of National Relevance (PRIN), project code 2020SKJSTF (“At the frontier of agent-based modelling: a new data driven framework for policy design toward sustainable and resilient economies”).

%Bibliography
\bibliographystyle{unsrt}  
\bibliography{references}  

\newpage
\section*{Appendix}\label{sec5}

\subsection*{Appendix A. A detailed break down on the correlations between disruptions and citations: additional details}\label{sec51}

\begin{figure}[h!]
\centering
\includegraphics[width=0.7\textwidth]{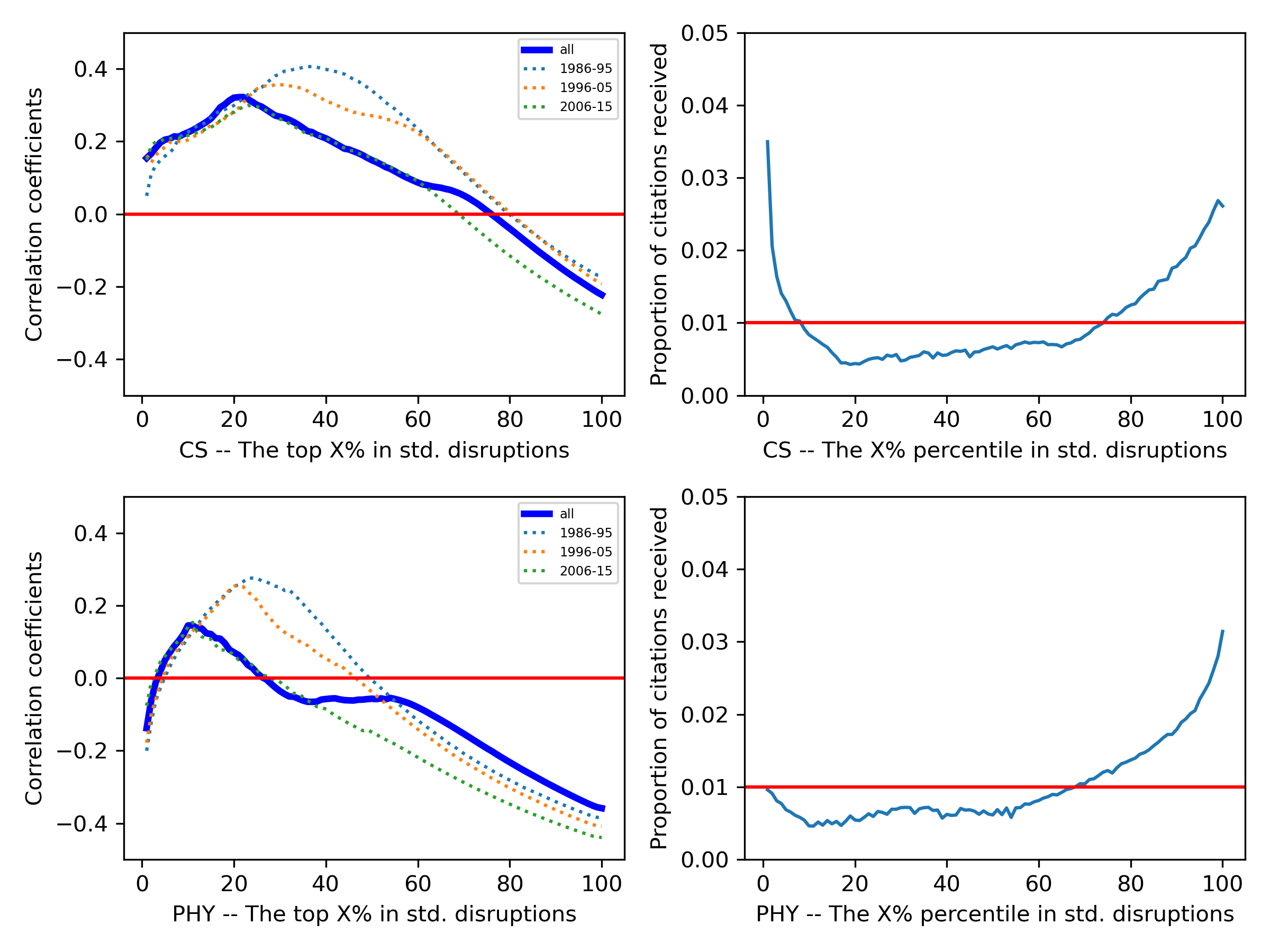}
\caption{The robustness check for Fig.~\ref{fig:R1Fig} under the standardized disruption score. Left column: Correlation coefficients across different percentiles of top disruptive papers as measured in standardized disruption scores in Computer Science (top) and Physics (bottom). Right column: The proportion of citations received by each percentile of Computer Science (top) and Physics (bottom) papers.}
\label{fig:R1Supp1}
\end{figure}

\begin{figure}[h!]
\centering
\includegraphics[width=0.7\textwidth]{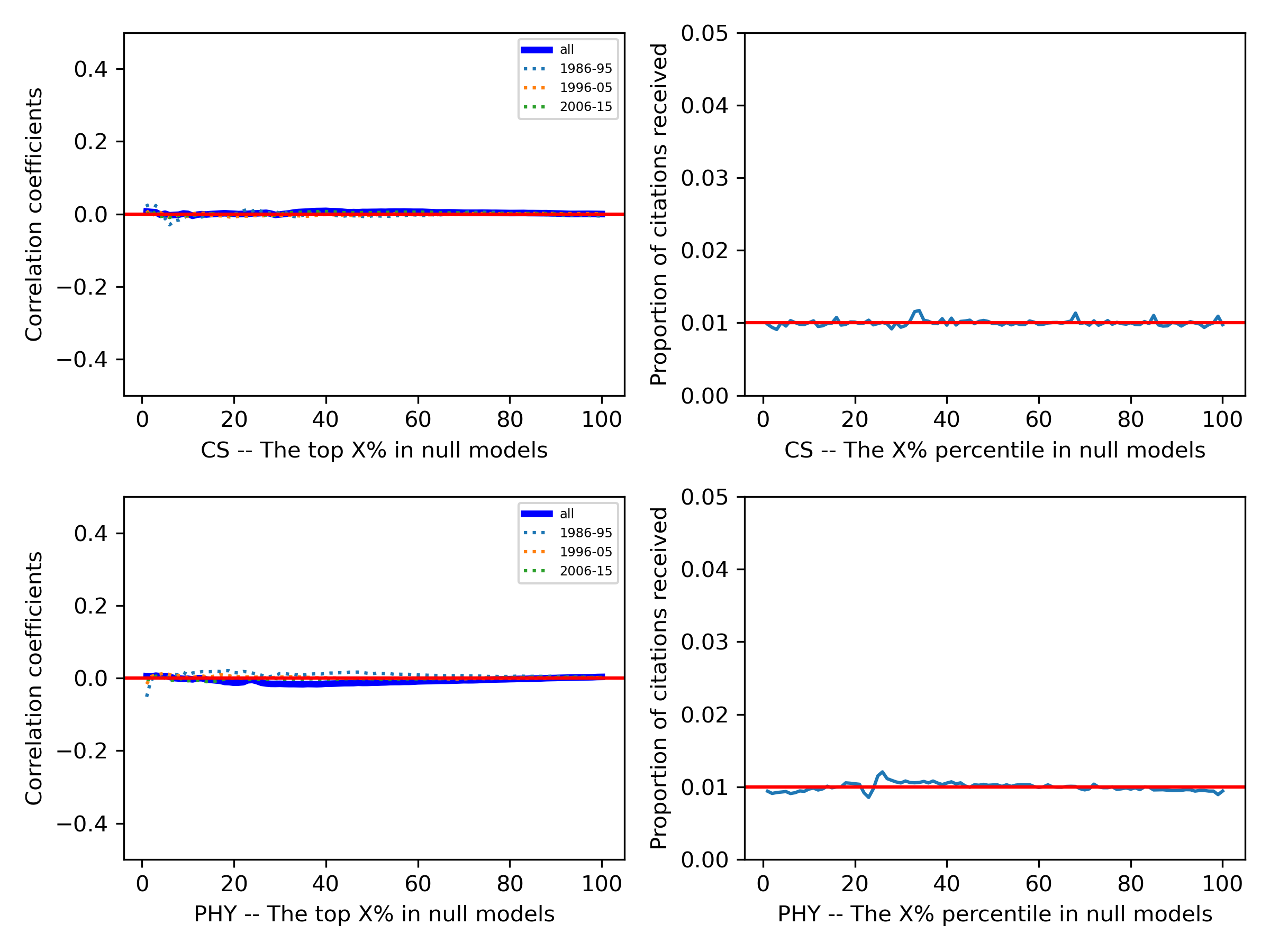}
\caption{The robustness check for Fig.~\ref{fig:R1Fig} under the null model. We create the null model by reshuffling the 5-year accumulated citations received by each paper. Under the null model, the left column represents correlation coefficients across different percentiles of top disruptive papers in Computer Science (top) and Physics (bottom), and the right column presents the proportion of citations received by each percentile of Computer Science (top) and Physics (bottom) papers.}
\label{fig:R1Supp2}
\end{figure}

We investigate the correlation coefficients between disruption scores and citation counts across different percentiles of top disruptive papers. Our results are obtained using the original disruption score, which is calculated directly from its defining formula. To further validate our results, we firstly run the experiments as described in~\ref{sec21} with standardized disruption scores. Then we repeat the experiment with randomised null models. The outcomes are illustrated in Fig.~\ref{fig:R1Supp1} and Fig.~\ref{fig:R1Supp2}, respectively. As illustrated in these figures, the correlation trajectories in both Computer Science and Physics remain consistent with our main results under the standardized disruption metric, and the observed correlation patterns cannot be explained by the null models.

\subsection*{Appendix B. Most-cited papers are relatively less disruptive: additional details}\label{sec52}

\begin{figure}[h!]
\centering
\includegraphics[width=0.7\textwidth]{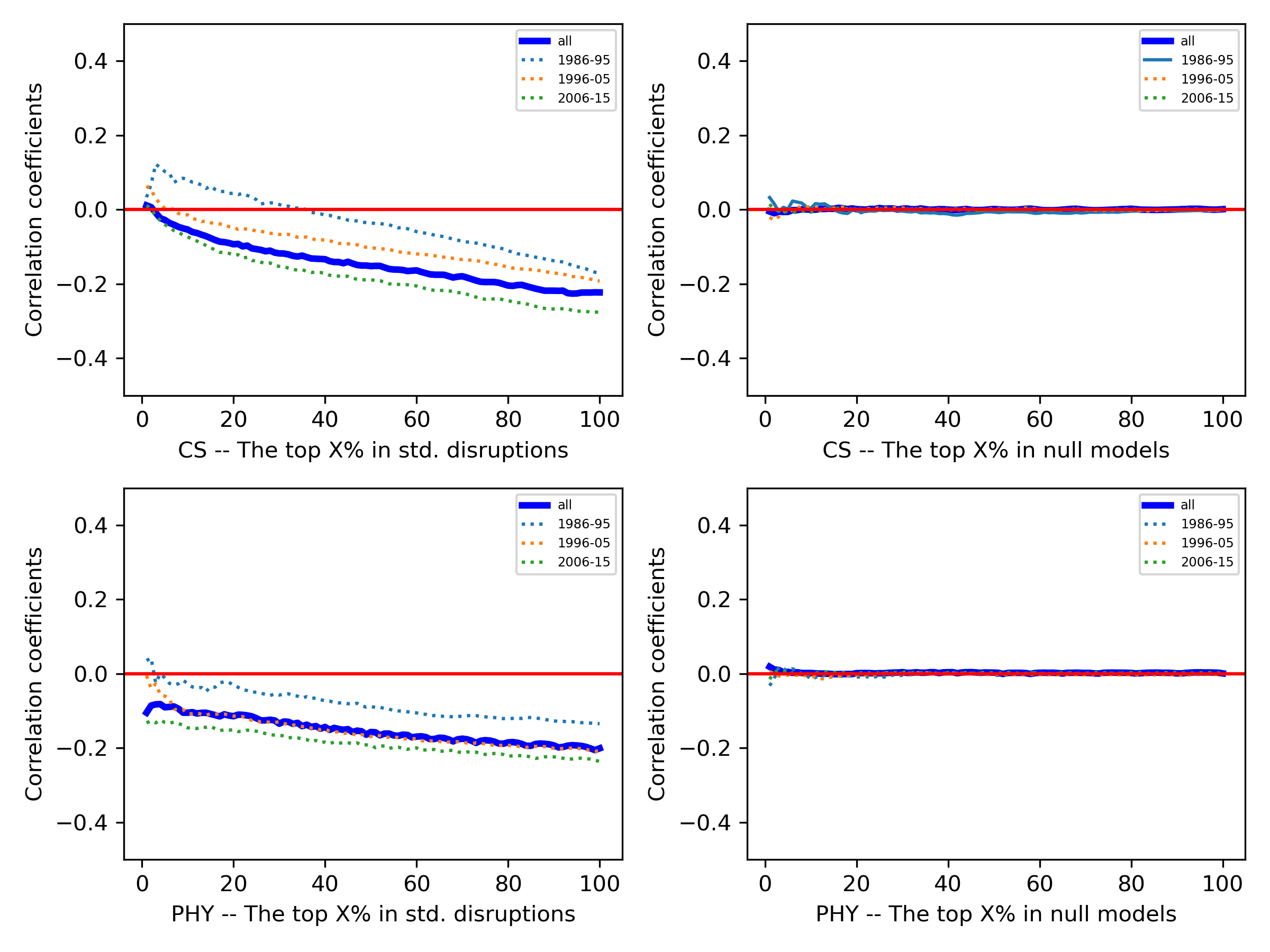}
\caption{The robustness check for Fig.~\ref{fig:R2Fig}. Left column: Correlation coefficients across different percentiles of most-cited papers in Computer Science (top) and Physics (bottom) under the standardized disruption score. Right column: Correlation coefficients across different percentiles of most-cited papers in Computer Science (top) and Physics (bottom) under the null model.}
\label{fig:R2Supp1}
\end{figure}

In this section, we provide robustness checks for our results on the correlations between disruptions and citations across various percentiles of most-cited papers. We first replicate the main results with standardized disruption scores (see the left column of Fig.~\ref{fig:R2Supp1}). One should note that standardizing the disruption score changes the order of papers within the disruption rank, which can lead to different correlation coefficients. In addition, we run the experiment with null models (see the right column of Fig.~\ref{fig:R2Supp1}) as well. Here the null models are constructed by reshuffling the disruption scores achieved by papers in our datasets. Again, we find that our main results in~\ref{sec22} still hold with the standardized scores in both disciplines, and our results are significantly different from the outcomes derived from the null models.

\subsection*{Appendix C. The relationship between disruptions and citations in scientific careers: additional details}

\begin{figure}[h!]
\centering
\includegraphics[width=0.7\textwidth]{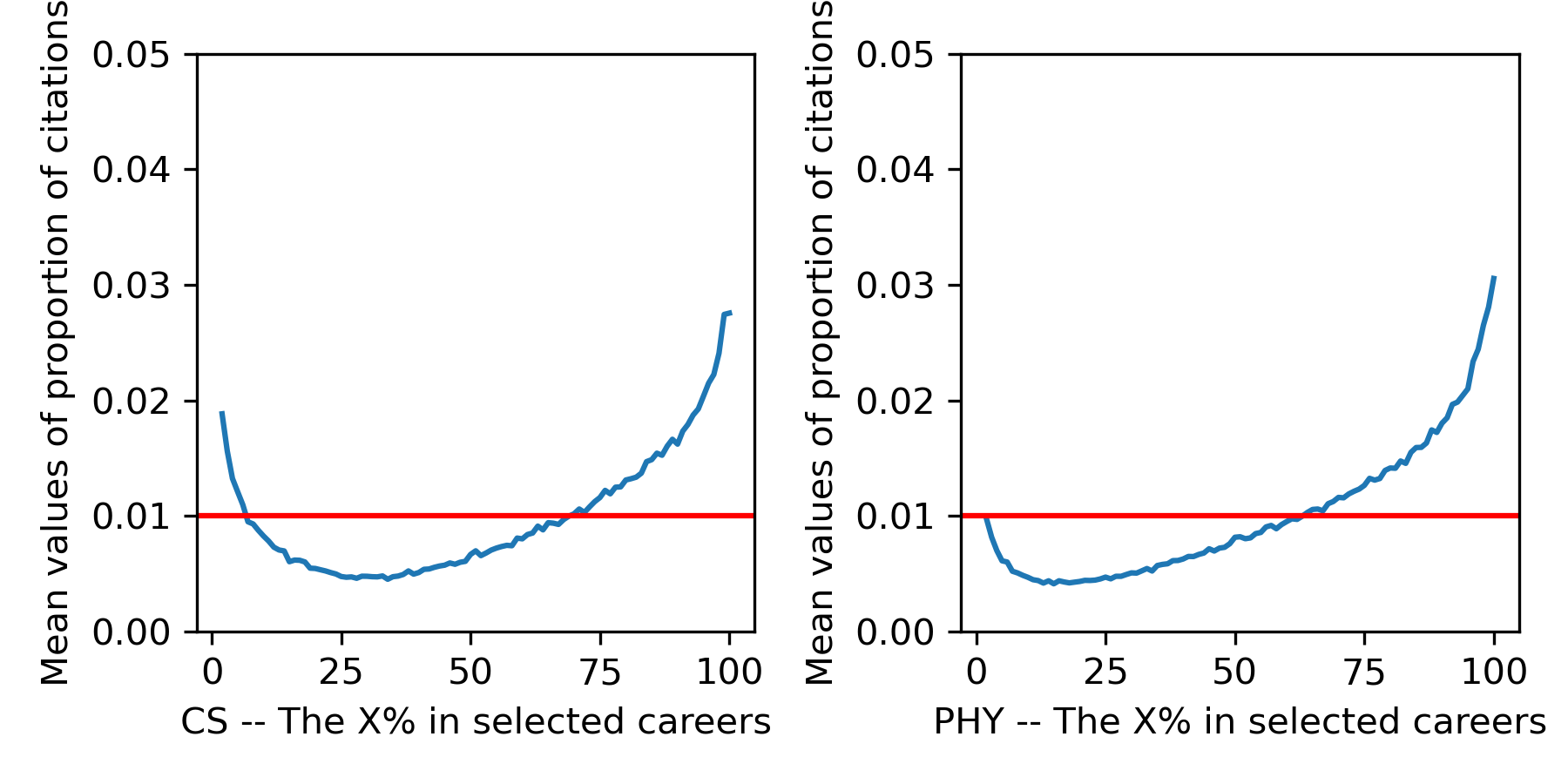}
\caption{Mean values of proportion of received citations for each percentile of disruptive papers in the careers of Computer Science (left) and Physics (right). Here the figures are constructed based on researchers who have more than 100 publications. We can see that these figures are similar to the curves in (b) and (d) of Fig.~\ref{fig:R1Fig}.}
\label{fig:R3Supp3}
\end{figure}

In the main text, we demonstrate that the paper-level results on the relationship between disruptions and citations remain equivalent in scientific careers. The major difference in the career settings is that in Fig.~\ref{fig:R3Fig1} (b) and (d), the values of the curves are inflated when compared to the results from the paper datasets. To explain such a phenomenon, we illustrate the proportion of citation curves for researchers who have published more than 100 papers in Fig.~\ref{fig:R3Supp3}. The new curves are verisimilar to our paper-level results.

\begin{figure}[h!]
\centering
\includegraphics[width=0.7\textwidth]{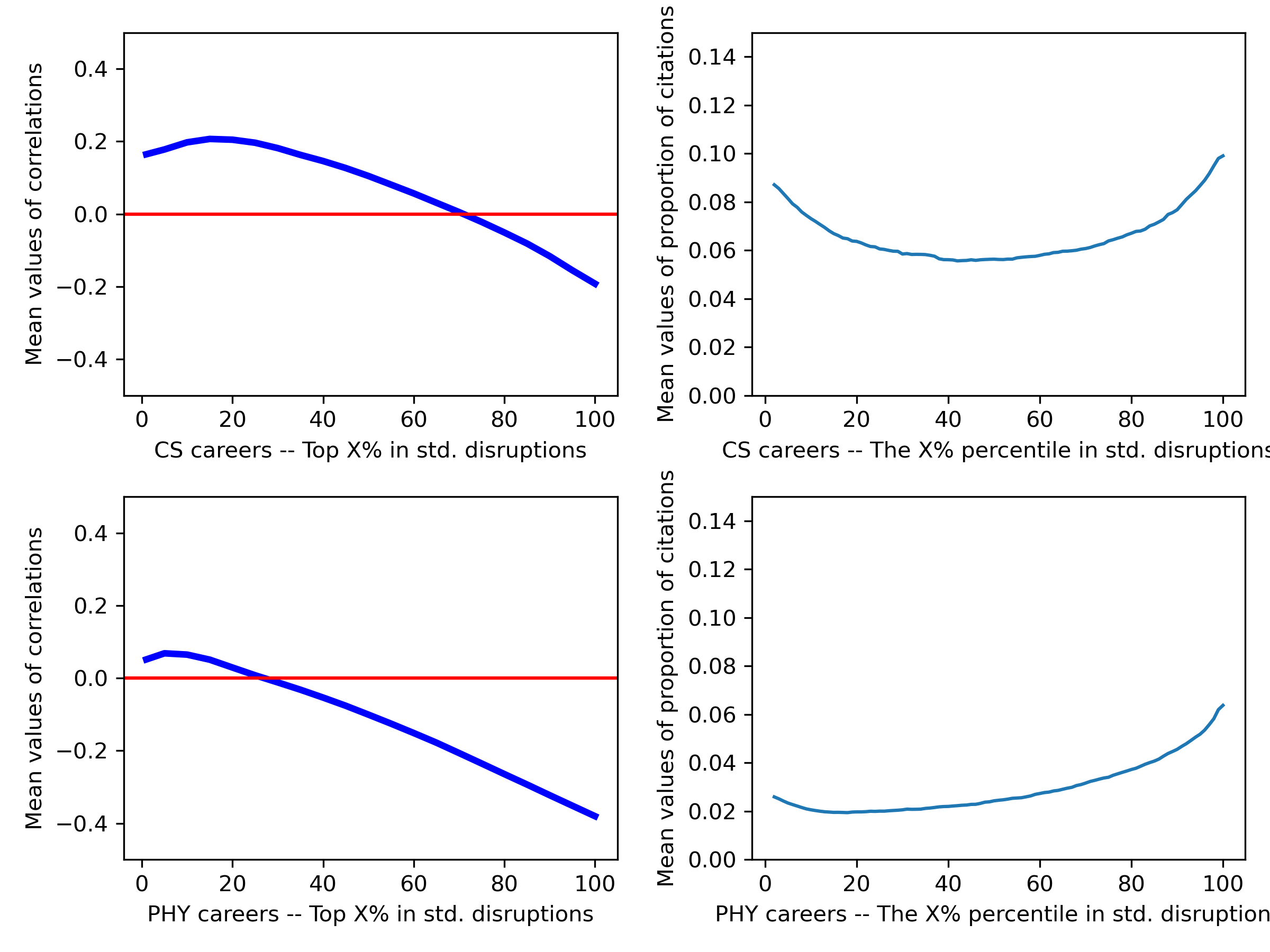}
\caption{The robustness check for Fig.~\ref{fig:R3Fig1} under the standardized disruption score. Left column: mean values of correlation coefficients across different percentiles of top disruptive papers as measured in standardized disruption scores in the careers of Computer Scientists (top) and Physicists (bottom). Right column: mean values of proportion of citations received by each percentile of papers in the publication profile of Computer Scientists (top) and Physicists (bottom).}
\label{fig:R3Supp1}
\end{figure}

\begin{figure}[h!]
\centering
\includegraphics[width=0.7\textwidth]{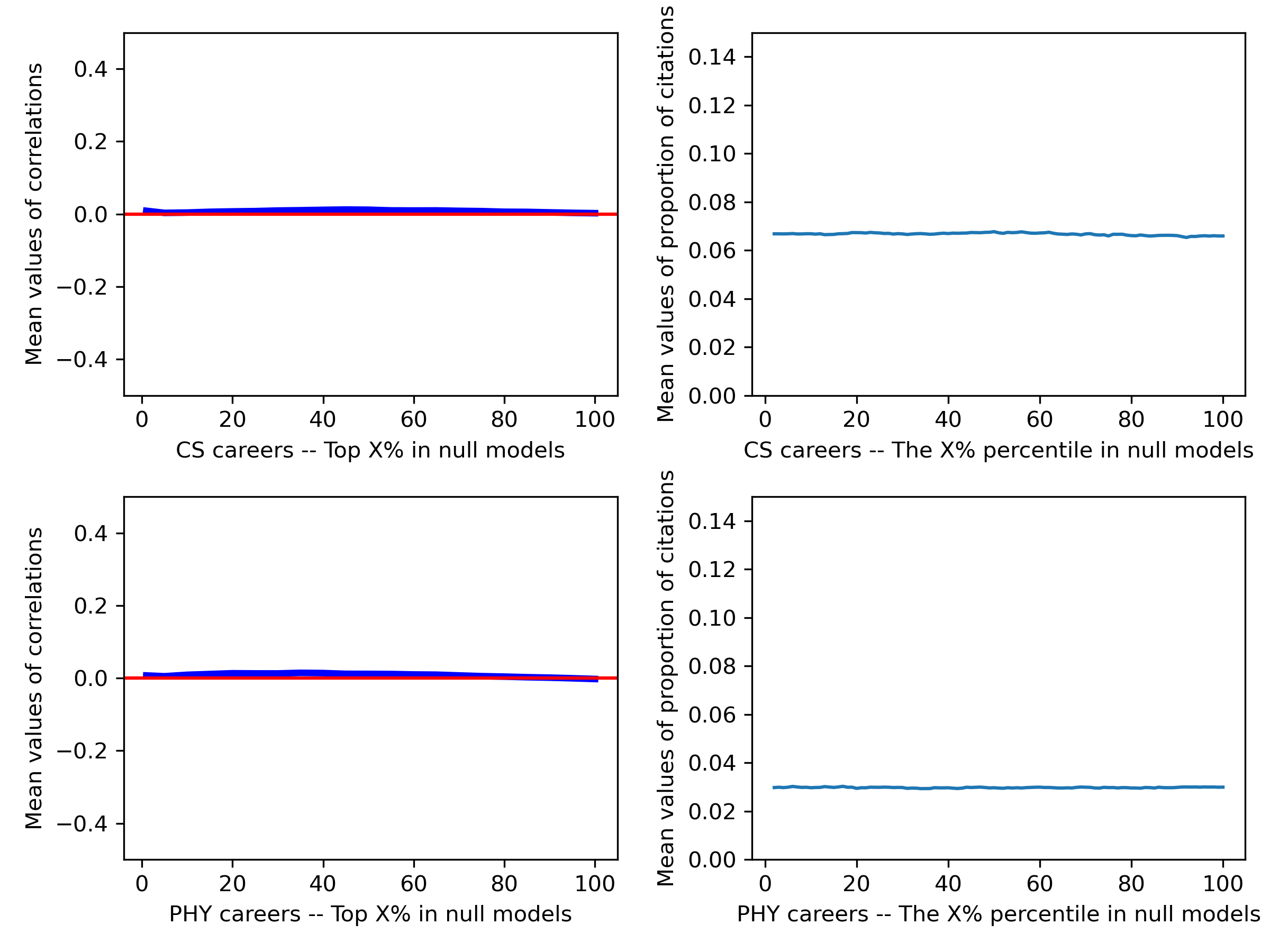}
\caption{The robustness check for Fig.~\ref{fig:R3Fig1} under the null model. Left column: mean values of correlation coefficients across different percentiles of top disruptive papers in the careers of Computer Scientists (top) and Physicists (bottom). Right column: mean values of proportion of citations received by each percentile of papers in the publication profile of Computer Scientists (top) and Physicists (bottom). It is apparent that our results in Fig.~\ref{fig:R3Fig1} cannot be explained by the null models.}
\label{fig:R3Supp2}
\end{figure}

\begin{figure}[h!]
\centering
\includegraphics[width=0.7\textwidth]{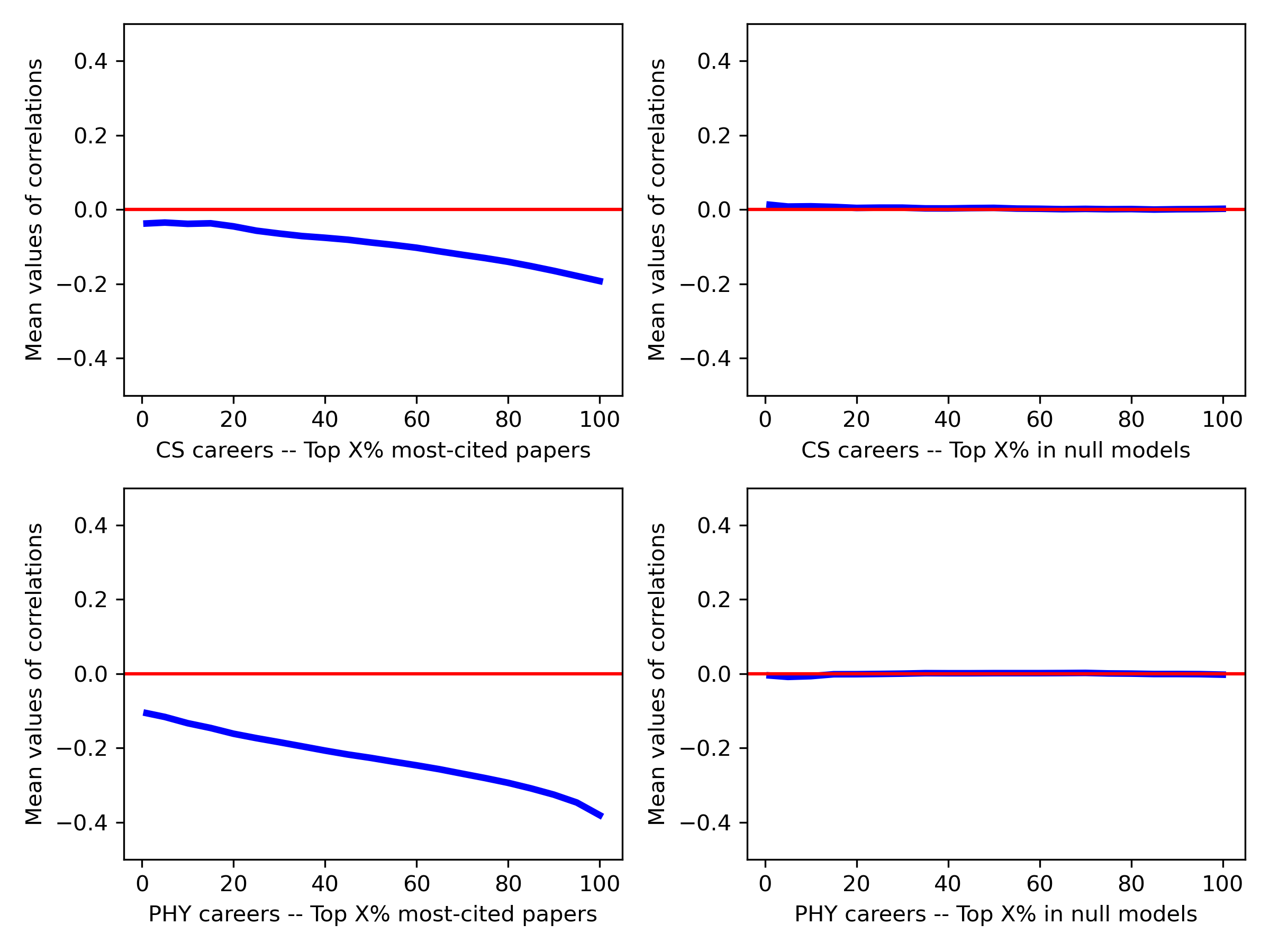}
\caption{The robustness check for Fig.~\ref{fig:R3Fig2}. Left column: mean values of correlation coefficients across different percentiles of most-cited papers papers in the careers of Computer Scientists (top) and Physicists (bottom). Here the correlation coefficients are computed based on standardized disruption scores. Right column: mean values of correlation coefficients across different percentiles of most-cited papers in Computer Science (top) and Physics (bottom) under the null model.}
\label{fig:R3Supp4}
\end{figure}

Moreover, we further corroborate our career-level results by adopting standardized disruption scores and by comparing our original results to the results of the null model. According to Fig.~\ref{fig:R3Supp1}, Fig.~\ref{fig:R3Supp2}, and Fig~\ref{fig:R3Supp4}, we find that our career-level results still hold with standardized disruption scores and that these results are robust against null models.

\newpage
% R1 std. disruption

% R1 null model

% R2 std. score and null model

% R3 career-analysis, 100+ papers researchers

% R3 career-analysis of R1, std. scores for top disruptive papers 

% R3 career-analysis of R1, null models 

% R3 career-analysis of R2, std. scores and null models

\end{document}